\documentclass{Interspeech2024}

\usepackage{cite}
\usepackage{url}
\usepackage{makecell, hhline}
\usepackage{float}
\usepackage{amsmath,amssymb,amsfonts}
\usepackage{algorithmic}
\usepackage{graphicx}
\usepackage{textcomp}
\usepackage{xcolor}
\usepackage[nolist,nohyperlinks]{acronym}
\usepackage{multirow}
\usepackage[subscriptcorrection,varg,varvw,cmbraces,cmintegrals]{newtxmath}
\usepackage{caption, subcaption} 
\interspeechcameraready 

\title{Self-Supervised Embeddings for Detecting Individual Symptoms of Depression}

\name[affiliation={1,2}]{Sri Harsha}{Dumpala}
\name[affiliation={3}]{Katerina}{Dikaios}
\name[affiliation={1}]{Abraham}{Nunes}
\name[affiliation={1,2}]{Frank}{Rudzicz}
\name[affiliation={1,4}]{Rudolf}{Uher}
\name[affiliation={1,2}]{Sageev}{Oore}

\address{
  \{$^1$Dalhousie University, $^2$Vector Institute, $^3$McMaster University, $^4$Nova Scotia Health\}, Canada}

\email{\{sriharsha.d, nunes, frank, uher, sageev\}@dal.ca, dikaiosk@mcmaster.ca} 


\begin{document}

\maketitle
\begin{abstract}
Depression, a prevalent mental health disorder impacting millions globally, demands reliable assessment systems. Unlike previous studies that focus solely on either detecting depression or predicting its severity, our work identifies individual symptoms of depression while also predicting its severity using speech input. We leverage self-supervised learning (SSL)-based speech models to better utilize the small-sized datasets that are frequently encountered in this task. Our study demonstrates notable performance improvements by utilizing SSL embeddings compared to conventional speech features. We compare various types of SSL pretrained models to elucidate the type of speech information (semantic, speaker, or prosodic) that contributes the most in identifying different symptoms. Additionally, we evaluate the impact of combining multiple SSL embeddings on performance. Furthermore, we show the significance of multi-task learning for identifying depressive symptoms effectively.
\end{abstract}

\section{Introduction}
\label{intro}

Major depressive disorder, commonly known as depression, is the most prevalent mental health disorder, and a leading cause of global disability~\cite{rehm2019global}. Yet, it remains largely under-detected and under-treated~\cite{herrman2022time}.  
Therefore, accurately identifying and treating depression is a significant public health priority aimed at reducing its prevalence. However, scarcity of trained clinicians and other resources hinder this objective. While deep learning-based mental health assessment systems have shown promise in tasks including patient screening and monitoring, further research is necessary to develop more informative and interpretable deep learning models for widespread deployment in clinical settings.

Recent research has demonstrated the efficacy of utilizing natural speech as a potentially inexpensive and easily scalable indicator for assessing depression~\cite{cummins2015review, avec_2019, yamamoto2020using, low2020automated, dikaios2023applications}. Previous studies primarily focused on either detecting depression \cite{harati2021speech, dubagunta2019learning, ravi2024enhancing} or predicting its severity \cite{dumpala2021estimating, seneviratne2022multimodal, dumpala2021sine, nature_sc_reports_2023}. Potential vocal biomarkers for depression identified in these studies encompass a spectrum of traditional acoustic features, including prosodic elements (such as pitch and speech rate), spectral characteristics (like Mel-frequency cepstral coefficients and formant frequencies), and glottal attributes (related to vocal fold behaviour)~\cite{avec_2019, dubagunta2019learning, low2020automated}, as well as neural embeddings from pretrained models like $x$-vectors, wav2vec 2.0, and HuBERT~\cite{nature_sc_reports_2023, ssl_dep_2023, combining_interspeech}. However, these studies often operate under the assumption that of all the measurable aspects of speech, depression severity is the most significant and influential \cite{fara2023bayesian}. Yet, merely identifying or estimating depression severity may prove inadequate compared to the clinical setting of monitoring the core subset of depressive symptoms for effective treatment outcomes \cite{txt_ind_sym_2020}. In this study, we adopt a clinically oriented approach to depression assessment -- aiming to detect depressive symptoms while also predicting overall depression severity. This method also offers an implicit explanation for the decisions made by the automated models.

Only a few studies have explored the task of identifying individual symptoms of depression using speech as input~\cite{FaraGMC22, takano2023estimating, fara2023bayesian}. Fara et al.~\cite{FaraGMC22, fara2023bayesian} used acoustic features such as eGeMAPS \cite{eyben2015egemaps} and prosodic features extracted from spontaneous speech (in English) to train random forest classifiers for detecting individual symptoms of depression based on the self-reported Patient Health Questionnaire (PHQ-8)~\cite{phq8-paper}. Similarly, Takano et al.~\cite{takano2023estimating} used spectral, prosodic, and glottal features extracted from read speech (in Japanese) to train decision tree models for detecting individual symptoms of depression based on the clinician-rated HAM-D scale \cite{williams2008grid}. These works used conventional speech features to train simple machine learning models for detecting individual symptoms of depression. In this work, we use neural embeddings extracted from deep learning models to detect individual symptoms of depression based on the Montgomery and Åsberg Depression Rating Scale (MADRS) \cite{madrs_scale}, which is considered a gold standard for clinical assessment of depression severity.

One of the major hindrances in applying deep learning techniques to the medical domain, including depression assessment, is insufficient training data \cite{rutowski2022toward}. To address the challenge of low-resourced data, we leverage SSL-based training to exploit the availability of large amounts of unlabeled speech data. Previous studies have employed SSL-based speech models to detect depression \cite{zhang2021depa, campbell2023classifying, ssl_dep_2023}. We extend these approaches by leveraging SSL-based speech models to both (A) identify individual symptoms of depression and (B) predict depression severity. Specifically, we analyze several SSL-based speech models pretrained with different objective functions such as masked language modeling (MLM), contrastive learning, and speaker-invariant training ~\cite{hsu2021hubert, 2023byol-a, Contentvec}. The choice of the pre-training objective function defines the information encoded by the SSL speech models, such as semantic, speaker, and prosodic information. Using these models, we assess the significance of each information type in identifying individual symptoms of depression. Additionally, we examine the impact of combining embeddings obtained from different SSL models in identifying individual symptoms of depression. 

\begin{figure}[tb]
\centering
 \begin{subfigure}{0.43\linewidth}
         \centering
         \includegraphics[width=0.93\linewidth]{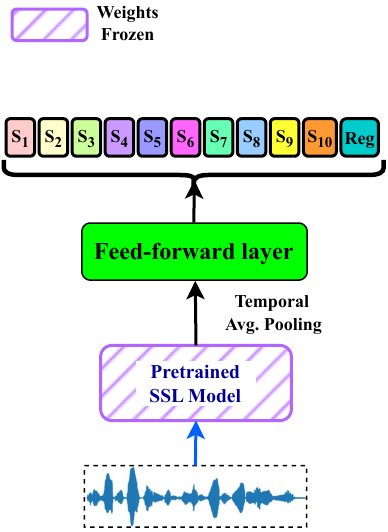}
         \caption{Single SSL model}
         \label{fig:single_ssl}
     \end{subfigure}
\begin{subfigure}{.55\linewidth}
         \centering
         \includegraphics[width=0.97\linewidth]{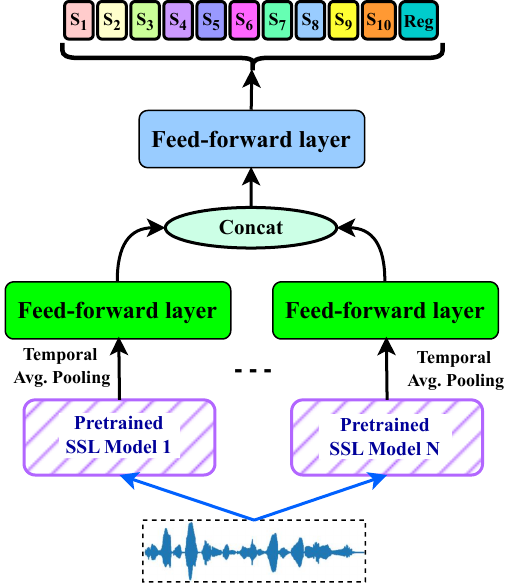}
         \caption{Combining multiple SSL models}
         \label{fig:ccomb_ssl}
\end{subfigure}
\caption{Schematic diagram of the symptom detection model using (a) single and (b) multiple SSL models (N models). In this work, N = 2 or 3. Multi-task setting will have 10 symptom detection heads in the output layer ($S_i$ for $1 \leq i \leq 10$) along with one regression (Reg) head. In single-task setting, there will be only one head (specific-symptom detection or regression).}
\label{fig:block_diag}

\end{figure}

\section{SSL-models for Analysis}
In this paper, we analyze several SSL-based speech models pretrained using different objective functions. The type of speech information, whether semantic, speaker, or prosodic, encoded by the SSL model depends on the pre-training objective function. Different pre-training objectives used to train SSL-based speech models include masked language modeling (MLM), contrastive pre-training, speaker-invariant training, and student-teacher training with cross-entropy loss~\cite{hsu2021hubert, 2023byol-a, Contentvec, rdino_2023}. Below, we explain the SSL models that we use in this paper. 
\begin{itemize}
\item HuBERT~\cite{hsu2021hubert} was pretrained using MLM which uses an offline clustering step to align target labels. This model predominantly encodes semantic information in speech.
\item WavLM~\cite{chen2022wavlm} was pretrained using MLM and denoising objectives. WavLM predominantly encodes semantics with some speaker-specific information. Here, we use the WavLM base model.
\item BEATS~\cite{beats} uses an iterative framework to optimize an acoustic tokenizer and an audio SSL model. The audio SSL model is optimized using MLM where the discrete labels are generated by the acoustic tokenizer. This objective allows BEATS to learn semantic-rich embeddings.
\item ContentVec~\cite{Contentvec} uses a speaker-invariant training objective to suppress speaker information contained in the pretrained HuBERT model. ContentVec encodes semantic information while discarding speaker-specific information.
\item RDINO \cite{rdino_2023} uses a regularized distillation with no labels (DINO) framework to pre-train models to encode speaker-specific information.
\item AudioMAE~\cite{AudioMAE} is an encoder-decoder model pretrained using MLM with a very high masking ratio -- around 80\% of frames masked. The pre-training task of reconstruction makes the model encode semantic, speaker, and prosodic information.
\item BYOL-Audio~\cite{2023byol-a} was pretrained by maximizing the cosine similarity between multiple augmented versions of the same speech signal. This model encodes the global information contained in speech, i.e., mainly speaker and prosodic. 
\end{itemize}

Note that WavLM, BEATS, HuBERT, ContentVec, and RDINO take raw speech as input whereas BYOL-Audio and AudioMAE take spectrograms. As shown in Figure~\ref{fig:block_diag}, we freeze all the weights of these pretrained models, and train only the feed-forward and output layers to detect the individual symptoms of depression. We analyze SSL models using two different architectures: (1) single SSL model -- here we use the embeddings of a single SSL model (Figure \ref{fig:single_ssl}), and (2) Multiple SSL models -- here we combine embeddings of multiple SSL models (Figure \ref{fig:ccomb_ssl}). Moreover, we evaluate each of this architecture in single-task and multi-task settings. For the multi-task setting, the output layer consists of $10$ classification heads, each with two softmax units to detect each symptom, and a single linear unit for predicting the overall depression severity. For the single-task setting, we train a separate model for each symptom which consists of a single classification head as the output layer.

We also use conventional speech features such as spectrograms, COVAREP~\cite{degottex2014covarep}, and the extended Geneva Minimalistic Acoustic Parameter Set (eGeMAPS)~\cite{eyben2015egemaps} as baselines. We extract 80-dimensional spectrograms, 88-dimensional eGeMAPS features using OpenSMILE~\cite{eyben2010opensmile}, and 74-dimensional COVAREP features using COVAREP~\cite{degottex2014covarep} to train convolutional neural network (CNN) models. The CNNs trained using conventional speech features consist of two convolutional layers, each with 100 channels. The kernels have sizes of 3 and 5 for the first and second layers, respectively. The outputs of the second layer are then passed through a fully connected layer with $100$ sigmoid linear units (SiLU) \cite{silu_activation}. Subsequently, a symptom-specific classification head with two softmax units follows the fully connected layer. Separate CNN models are trained for detecting each symptom. The output layer of the CNN model trained to predict overall depression severity consists of a linear unit.

\begin{table}[!htbp]
\caption{Individual symptoms of depression as defined in MADRS and their abbreviations, {\em Abbr.}} 
\label{tab:madrs_symptoms}
\centering
\vspace{0.2cm}
\resizebox{1.01\linewidth}{!}{%
\begin{tabular}{ll||ll} 
\toprule
Symptom & Abbr. & Symptom & Abbr. \\
\midrule
1) Apparent Sadness & ASad & 6) Concentration Difficulties & ConD \\
2) Reported Sadness & RSad & 7) Lassitude & Lass \\
3) Inner Tension & InTen & 8) Inability to Feel & IFeel \\
4) Reduced Sleep & RSlp & 9) Pessimistic Thoughts & PesT \\
5) Reduced Appetite & RApp & 10) Suicidal Thoughts & SuiT \\
\bottomrule
\end{tabular}}
\end{table}

\begin{figure}
  \centering
  \includegraphics[width=0.59\linewidth]{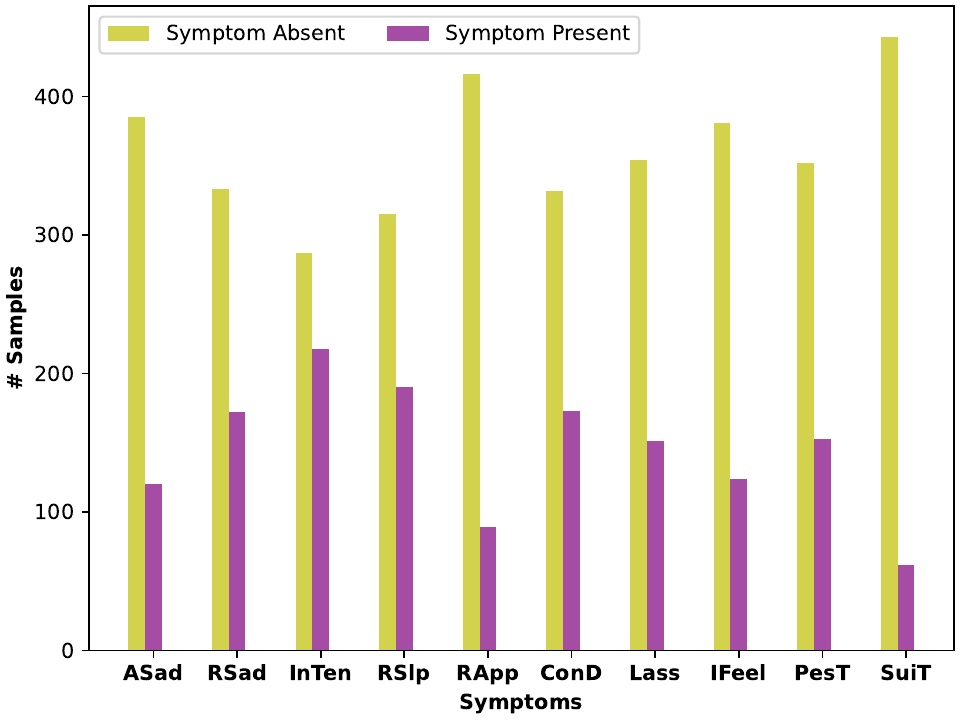}
  \caption{Distribution of samples between class-0 (symptom absent) and class-1 (symptom present) for each symptom in the MADRS. Symptom abbreviations are provided in Table \ref{tab:madrs_symptoms}.}
  \label{fig:dist_symptoms}
\end{figure}

\begin{table*}[!htbp]
\caption{\textbf{Performance of different self-supervised learning speech models in single task setting}: Comparison of the model performance (for each individual symptom of MADRS). Each cell contains F-scores in the form: macro F-score $F_{M}$ (Symptom absent $F_{A}$, Symptom present $F_{P}$). RMSE shown for the regression task in predicting the overall depression severity score according to MADRS. Best performance (in terms of $F_M$) for each symptom is presented in bold. a-priori refers to the system that always predicts the output as the majority class i.e., symptom absent.} 
\label{tab:ssl_single-task}
\centering
\vspace{0.2cm}
\resizebox{0.91\linewidth}{!}{%
\begin{tabular}{l|cccccccc} 
\toprule
Symptoms & a-priori & WavLM & HuBERT & BEATS & ContentVec & RDINO & BYOL-Audio & AudioMAE \\
\midrule
(1) ASad & 43 (86, 0) & \textbf{63 (84, 42)} & 60 (79, 41) & 59 (81, 36) & 57 (73, 41) & 62 (81, 43) & 57 (77, 37) & 62 (83, 41) \\
(2) RSad & 40 (79, 0) & \textbf{72 (79, 65)} & 71 (74, 68) & 68 (75, 61) & 61 (64, 57) & 69 (72, 66) & 68 (74, 61) & 68 (72, 64) \\
(3) InTen & 36 (72, 0) & 68 (71, 65) & 68 (70, 66) & 67 (69, 64) & 58 (57, 59) & \textbf{69 (70, 68)} & 64 (65, 63) & 66 (64, 67) \\
(4) RSlp & 38 (76, 0) & 66 (75, 57) & \textbf{67 (73, 61)} & 64 (73, 55) & 58 (61, 55) & 66 (73, 59) & 62 (64, 59) & 65 (73, 56) \\
(5) RApp & 45 (90, 0) & 47 (84, 10) & 48 (87, 08) & 52 (87, 18) & 41 (75, 07) & 47 (88, 06) & 42 (79, 05) & \textbf{53 (88, 17)}\\
(6) ConD & 40 (79, 0) & 70 (80, 59) & \textbf{72 (79, 65)} & \textbf{72 (82, 61)} & 65 (70, 61) & 67 (74, 59) & 69 (75, 63) & 69 (77, 61) \\
(7) Lass & 41 (82, 0) & \textbf{68 (77, 58)} & 66 (78, 54) & 61 (75, 47) & 57 (64, 51) & 64 (76, 52) & 59 (73, 45) & 65 (78, 52) \\
(8) IFeel & 43 (86, 0) & 68 (76, 59) & 68 (77, 60) & 63 (79, 46) & 60 (63, 57) & \textbf{73 (81, 64)} & 61 (77, 45) & 66 (77, 55) \\
(9) PesT & 41 (82, 0) & 72 (80, 64) & 72 (75, 70) & 69 (73, 64) & 65 (68, 62) & 73 (70, 75) & \textbf{74 (81, 66)} & 71 (73, 68) \\
(10) SuiT & 47 (93, 0) & 59 (82, 36) & 60 (83, 36) & 54 (79, 29) & 55 (79, 31) & \textbf{63 (82, 43)} & 60 (83, 37) & 61 (82, 39) \\
\midrule
MADRS(RMSE) & -- & 8.53 & \textbf{8.44} & 8.92 & 9.46 & 8.76 & 8.85 & 9.05 \\
\bottomrule
\end{tabular}}
\end{table*}

\section{Dataset Details}
We use an internal speech-based depression dataset which we call the Clinically-Labelled with INdividual symptoms of Depression (CLIND) dataset. CLIND consists of speech samples in English language collected from $505$ participants ($351$ female and $154$ male). Each participant was asked to speak about their experiences from the past few weeks with different prompts. The three prompts were designed to evoke neutral, positive, and negative contexts, respectively. The interviewer let the participant speak uninterrupted and only used a standardized prompt if the participant was silent for 30 seconds. At least $4$ minutes of speech, including pauses, were recorded following each prompt, providing a total of approximately 10 minutes of speech per participant. Trained clinicians used structured interviews to score each participant's current depression severity on MADRS, which consists of 10 individual symptoms (see Table \ref{tab:madrs_symptoms}), with each symptom scored in the range of $0-6$, and a total score in the range of $0-60$. The range of total MADRS scores in our dataset range from $0$ to $47$. In this study, we consider a symptom absent (Class-0) if the score for a symptom ($S_s$) is $0$ $\leq$ $S_s$ $\leq$ $1$, and a symptom to be present (Class-1) if $2$ $\leq$ $S_s$ $\leq$ $6$. The distribution of class-level samples for each symptom is illustrated in Figure \ref{fig:dist_symptoms}. To train and test the models, we segment the speech samples into $10$-second segments, resulting in a total of $25,471$ segments (we segment after segregating the speech samples into train and test sets to avoid speaker overlap). We use the sample-level labels, encompassing both symptoms and overall depression severity scores, for all segments. Test performance is reported by majority voting across all segments of the test samples.

It is essential to note that the majority of publicly available depression datasets were labeled using self-reported scores like the Patient Health Questionnaire (PHQ-8)~\cite{phq8-paper}. In contrast, our dataset is rated by trained clinicians using the MADRS, which is regarded as a gold standard for clinical assessment of depression severity.

\begin{table}[!htbp]
\caption{\small Comparing performance (in terms of $F_M$ ($F_A, F_P$)) of conventional speech features to HuBERT (SSL-based model). `Spectro' refers to spectrograms.}
\label{tab:conv_sll_feats_single}
\centering
\vspace{0.2cm}
\resizebox{1.0\linewidth}{!}{%
\begin{tabular}{l|cccc} 
\toprule
Symptoms & Spectro & COVAREP & eGeMAPS & HuBERT \\
\midrule
(1) ASad & 47(51, 43) & 52(61, 43) & 53(63, 43) & \textbf{60(79, 41)} \\
(2) RSad & 60(63, 56) & 58(64, 52) & 59(65, 52) & \textbf{71(74, 68)} \\
(3) InTen & 54(60, 48) & 55(64, 46) & 58(61, 54) & \textbf{68(70, 66)} \\
(4) RSlp & 52(54, 50) & 57(65, 49) & 55(58, 52) & \textbf{67(73, 61)} \\
(5) RApp & 42(72, 11) & 41(77, 5) & 42(74, 10) & \textbf{48(87, 08)}\\
(6) ConD & 48(55, 41) & 53(62, 44) & 61(72, 50) & \textbf{72(79, 65)} \\
(7) Lass & 46(53, 39) & 48(59, 37) & 50(66, 34) & \textbf{66(78, 54)} \\
(8) IFeel & 56(66, 46) & 53(70, 36) & 54(72, 37) & \textbf{68(77, 60)} \\
(9) PesT & 61(67, 54) & 60(68, 52) & 62(66, 59) & \textbf{72(75, 70)} \\
(10) SuiT & 52(72, 32) & 49(77, 21) & 53(73, 32) & \textbf{60(83, 36)} \\
\midrule
MADRS & \multirow{2}{*}{10.81} & \multirow{2}{*}{10.18} & \multirow{2}{*}{9.82} & \multirow{2}{*}{\textbf{8.44}} \\
(RMSE) & & & & \\
\bottomrule
\end{tabular}}
\end{table}

\begin{table*}[!htb]
\caption{\textbf{Comparing single-task (ST) and multi-task (MT) learning}: Performance of the model (in terms of macro F-score) for each individual item of MADRS. MADRS (RMSE) refers to the RMSE error in estimating total depression severity based on MADRS score.} 
\label{tab:ssl_single_multi-task}
\centering
\hspace{-0.2cm}
\vspace{0.2cm}
\resizebox{0.97\textwidth}{!}{%
\begin{tabular}{l|cl|cl|cl|cl|cl|cl|cl|cl|cl} 
\toprule
Symptoms & \multicolumn{2}{|c|}{WavLM} & \multicolumn{2}{|c|}{HuBERT} & \multicolumn{2}{|c|}{BEATS} & \multicolumn{2}{|c|}{ContentVec} & \multicolumn{2}{|c|}{RDINO} & \multicolumn{2}{|c|}{BYOL-Audio} & \multicolumn{2}{|c}{AudioMAE} &\multicolumn{2}{|c|}{\{CV,RD,BY\}} & \multicolumn{2}{|c}{\{HB,RD,BY\}} \\
\midrule
& ST & MT & ST & MT & ST & MT & ST & MT & ST & MT & ST & MT & ST & MT & ST & MT & ST & MT \\
\midrule
(1) ASad & 63 & 65($\uparrow$) & 60 & 59($\downarrow$) & 59 & 58($\downarrow$) & 57 & 57 & 62 & 64($\uparrow$) & 57 & 57 & 62 & 63($\uparrow$) & 62 & 63($\uparrow$) & 65 & \textbf{66}($\uparrow$) \\
(2) RSad  & 72 & 72 & 71 & 71 & 68 & 70($\uparrow$) & 61 & 62($\uparrow$) & 69 & 70($\uparrow$) & 68 & 70($\uparrow$) & 68 & 69($\uparrow$)  & 69 & 71($\uparrow$) & 73 & \textbf{74}($\uparrow$) \\
(3) InTen & 68 & 68 & 68 & 67($\downarrow$) & 67 & 66($\downarrow$) & 58 & 56($\downarrow$) & 69 & \textbf{71}($\uparrow$) & 64 & 65($\uparrow$) & 66 & 64($\downarrow$)  & 68 & 67($\downarrow$) & 70 & 70 \\
(4) RSlp  & 66 & 66 & 67 & 67 & 64 & 65($\uparrow$) & 58 & 59($\uparrow$) & 66 & 66 & 62 & 62 &65 & 66($\uparrow$)  & 67 & 67 & \textbf{70} & 69($\downarrow$) \\
(5) RApp  & 47 & 45($\downarrow$) & 48 & 47($\downarrow$) & 52 & \textbf{53}($\uparrow$) & 41 & 42($\uparrow$) & 47 & 47  & 42 & 43($\uparrow$) & \textbf{53} & 50($\downarrow$)  & 49 & 48($\downarrow$) & \textbf{53} & 52($\downarrow$) \\
(6) ConD  & 70 & 71($\uparrow$) & 72 & 73($\uparrow$) & 72 & 72 & 65 & 62($\downarrow$) & 67 & 66($\downarrow$) & 69 & 70($\uparrow$) & 69 & 71($\uparrow$) & 72 & 72 & 74 & \textbf{75}($\uparrow$) \\
(7) Lass & 68 & 69($\uparrow$) & 66 & 67($\uparrow$) & 61 & 63($\uparrow$) & 57 & 60($\uparrow$) & 64 & 67($\uparrow$)  & 59 & 61($\uparrow$) &65 & 66($\uparrow$)  & 64 & 65($\uparrow$) & 68 & \textbf{70}($\uparrow$) \\
(8) IFeel & 68 & 69($\uparrow$) & 68 & 70($\uparrow$) & 63 & 61($\downarrow$) & 60 & 62($\uparrow$) & 73 & \textbf{75}($\uparrow$) & 61 & 61 & 66 & 67($\uparrow$)  & 73 & 72($\downarrow$) & 73 & \textbf{75}($\uparrow$) \\
(9) PesT  & 72 & 74($\uparrow$) & 72 & 75($\uparrow$) & 69 & 71($\uparrow$) & 65 & 63($\downarrow$) & 73 & 75($\uparrow$) & 74 & \textbf{75}($\uparrow$) & 71 & 69($\downarrow$)  & 77 & 77 & \textbf{79} & \textbf{79} \\
(10) SuiT  & 59 & 62($\uparrow$) & 60 & 63($\uparrow$) & 54 & 59($\uparrow$) & 55 & 57($\uparrow$)  & 63 &65($\uparrow$) & 60 & 64($\uparrow$) & 61 & 63($\uparrow$) & 67 & 68($\uparrow$) & 70 & \textbf{71}($\uparrow$) \\
\midrule
MADRS & \multirow{2}{*}{8.5} & \multirow{2}{*}{7.7($\downarrow$)} & \multirow{2}{*}{8.4} & \multirow{2}{*}{7.9($\downarrow$)} & \multirow{2}{*}{8.9} & \multirow{2}{*}{8.2($\downarrow$)} & \multirow{2}{*}{9.5} & \multirow{2}{*}{8.9($\downarrow$)} & \multirow{2}{*}{8.8} & \multirow{2}{*}{8.1($\downarrow$)} & \multirow{2}{*}{8.9}  & \multirow{2}{*}{8.2($\downarrow$)} & \multirow{2}{*}{9.1} & \multirow{2}{*}{8.5($\downarrow$)}  & \multirow{2}{*}{8.1} & \multirow{2}{*}{7.8($\downarrow$)} & \multirow{2}{*}{7.9} & \multirow{2}{*}{\textbf{7.5}($\downarrow$)}  \\
(RMSE) & & & & & & & & & & & & & & & & & & \\
\bottomrule
\end{tabular}}
\end{table*}

\section{Experiments}
\subsection{Model training}
For each pretrained SSL model (see Figure \ref{fig:block_diag}), we freeze all its weights and extract embeddings. We perform average pooling on the embeddings across the temporal dimension before passing as input to a feed-forward layer comprising $100$ SiLU. In the scenario of combining multiple SSL models, we pass the output embedding of each model through a $100$-dimensional feed-forward layer with SiLU units before concatenating the embeddings. Subsequently, the concatenated output is passed through a $100$-dimensional feed-forward layer with SiLU activation, followed by an output layer. We train both the feed-forward and output layers using the Adam optimizer with a learning rate of $1e-3$, a weight decay of $1e-5$, and a batch size of $32$. Negative log-likelihood (NLL) and mean squared error (MSE) loss functions are utilized for training models on classification and regression tasks, respectively. All experiments are conducted using a single A40 GPU. Each model is trained for 5 epochs, and the results are reported using 5-fold cross-validation, where the folds are speaker independent (no overlap of speakers between folds). A randomly selected subset ($10\%$) of the training set is allocated as the validation set for hyperparameter tuning.

We train CNNs on the conventional speech features using Adam optimizer with $\beta_1=0.9$, $\beta_2=0.99$, an initial learning rate of $0.001$, weight decay of $1e-5$, and a batch size of $32$. Dropout rates of $0.3$ and $0.4$ were used for the convolutional and fully connected layers, respectively to avoid model overfitting. NLL and MSE loss functions were used to train models on classification and regression tasks, respectively.

\subsection{Results}
The performance of the models is evaluated in terms of F-scores. We provide three F-scores: F-score of the symptom being absent ($F_{A}$), F-score of the symptom being present ($F_{P}$), and the macro F-score ($F_{M}$), computed as $F_{M}$ = ($F_{A}$ + $F_{P}$)/2. We perform 5-fold cross validation, and provide the performance metrics as the mean across the 5-folds.

Table \ref{tab:ssl_single-task} provides the performance of different SSL models, in single-task setting, for each of the symptoms as defined in the MADRS. Models such as WavLM and HuBERT which encode semantic information along with some speaker and prosodic information achieved better performance on ASad, RSad, RSlp, ConD, and Lass symptoms. By contrast, RDINO, which encodes speaker information, achieved better performance on InTen, IFeel, and SuiT symptoms. Model BYOL-Audio, which encodes prosodic information, achieved better performance on PesT. ContentVec, which encodes semantic information and suppresses speaker information, achieved the lowest performance across all the models. These results suggest that semantic information alone, without speaker or prosodic information (as encoded by ContentVec), may not be sufficient to detect different symptoms of depression.

Table \ref{tab:conv_sll_feats_single} compares the performance of conventional speech features (Spectrogram, COVAREP and eGeMAP) with the embeddings extracted from SSL-based speech models. SSL-based models outperformed the conventional speech features. Among the conventional features, eGeMAPS perform better than Spectrogram and COVAREP features for most symptoms. Tables \ref{tab:ssl_single-task} and \ref{tab:conv_sll_feats_single} show that most of the SSL-based models outperform the conventional speech features across all symptoms. For the regression task, SSL models perform better than the conventional speech features with HuBERT achieving the best performance.

Table \ref{tab:ssl_single_multi-task} compares performance between single-task and multi-task learning. The systems trained with multi-task learning perform comparably or better than the single task learning systems. Moreover, the computational load of multi-task learning is much lower than the single-task learning system.
For all the models, the performance on the RApp (Reduced Appetite) symptom is very low. This could be due to the least correlation between the symptom RApp and speech \cite{uher2008measuring}.

Table \ref{tab:comb_sll_feats_single} provides the performance when different SSL-based speech embeddings are combined. These model combinations are selected such that different types of information (semantic, speaker and prosodic) can be comnbined. The performances are provided in terms of macro F-scores ($F_M$), and absolute improvement in performance when compared to the best performing single model in the set is provided in parenthesis. Combining semantic (ContentVec) with speaker (RDINO) or prosodic information (BYOL-Audio), i.e., \{CV,RD\} or \{CV,BY\}, improves performance. But combining semantic with speaker and prosodic information (\{CV,RD,BY\} and \{HB,RD,BY\}) further improves performance for most of the symptoms, with \{HB,RD,BY\} achieving the best performance on most of the symptoms. This shows that the semantic, speaker, and prosodic information might all be essential for most of the tasks.

\begin{table}[!htb]
\caption{Performance (in terms of macro F-score) obtained by combining different SSL-based speech models. CV, RD, BY and HB refer to ContentVec, RDINO, BYOL-A, HuBERT, respectively. The absolute improvement in performance (in terms of macro F-score) compared to the best performing single model in the set is given in parenthesis.}
\label{tab:comb_sll_feats_single}
\centering
\vspace{0.2cm}
\resizebox{1.0\linewidth}{!}{%
\begin{tabular}{l|lllll} 
\toprule
Symptoms & \{CV,RD\} & \{CV,BY\} & \{RD,BY\} & \{CV,RD,BY\} & \{HB,RD,BY\} \\
\midrule
(1) ASad & \hspace{0.1cm} 63 (1$\uparrow$) & \hspace{0.1cm} 61 (4$\uparrow$) & \hspace{0.1cm} 64 (2$\uparrow$) & \hspace{0.4cm} 62  & \hspace{0.34cm} \textbf{65} (3$\uparrow$) \\
(2) RSad & \hspace{0.1cm} 70 (1$\uparrow$) & \hspace{0.1cm} 68 & \hspace{0.1cm} 70 (1$\uparrow$) & \hspace{0.4cm} 69  & \hspace{0.34cm} \textbf{73} (2$\uparrow$)\\
(3) InTen & \hspace{0.1cm} 67 (2$\downarrow$)  & \hspace{0.1cm} 66 (2$\uparrow$) & \hspace{0.1cm} \textbf{70} (1$\uparrow$) & \hspace{0.4cm} 68 (1$\downarrow$) & \hspace{0.34cm} \textbf{70} (1$\uparrow$) \\
(4) RSlp & \hspace{0.1cm} 67 (1$\uparrow$) & \hspace{0.1cm} 63 (1$\uparrow$) & \hspace{0.1cm} 69 (3$\uparrow$) & \hspace{0.4cm} 67 (1$\uparrow$) & \hspace{0.34cm} \textbf{70} (3$\uparrow$) \\
(5) RApp & \hspace{0.1cm} 48 (1$\uparrow$) & \hspace{0.1cm} 44 (2$\uparrow$) & \hspace{0.1cm} 52 (5$\uparrow$) & \hspace{0.4cm} 49 (2$\uparrow$) & \hspace{0.34cm} \textbf{53} (5$\uparrow$) \\
(6) ConD & \hspace{0.1cm} 68 (1$\uparrow$) & \hspace{0.1cm} 72 (3$\uparrow$) & \hspace{0.1cm} 70 (1$\uparrow$) & \hspace{0.4cm} 72 (3$\uparrow$) & \hspace{0.34cm} \textbf{74} (2$\uparrow$) \\
(7) Lass & \hspace{0.1cm} 63 (1$\downarrow$) & \hspace{0.1cm} 61 (2$\uparrow$) & \hspace{0.1cm} 64  & \hspace{0.4cm} 64 & \hspace{0.34cm} \textbf{68} (2$\uparrow$) \\
(8) IFeel & \hspace{0.1cm} 72 (1$\downarrow$) & \hspace{0.1cm} 63 (2$\uparrow$) & \hspace{0.1cm} 72 (1$\downarrow$) & \hspace{0.4cm} \textbf{73}  & \hspace{0.34cm} \textbf{73} \\
(9) PesT & \hspace{0.1cm} 76 (3$\uparrow$) & \hspace{0.1cm} 75 (1$\uparrow$) & \hspace{0.1cm} 78 (4$\uparrow$) & \hspace{0.4cm} 77 (3$\uparrow$) & \hspace{0.34cm} \textbf{79} (5$\uparrow$) \\
(10) SuiT & \hspace{0.1cm} 66 (3$\uparrow$) & \hspace{0.1cm} 62 (2$\uparrow$) & \hspace{0.1cm} 68 (5$\uparrow$) & \hspace{0.4cm} 67 (4$\uparrow$) & \hspace{0.34cm} \textbf{70} (7$\uparrow$) \\
\midrule
MADRS & \hspace{0.23cm} 8.29 & \hspace{0.23cm} 8.35 & \hspace{0.23cm} 8.20 & \hspace{0.52cm} 8.12 & \hspace{0.5cm} \textbf{7.92}  \\
Score (RMSE) & \hspace{0.1cm} (0.47$\downarrow$) & \hspace{0.1cm} (0.50$\downarrow$) & \hspace{0.14cm} (0.56$\downarrow$) & \hspace{0.41cm} (0.64$\downarrow$) & \hspace{0.41cm} (0.52$\downarrow$) \\
\bottomrule
\end{tabular}}
\end{table}

We employ bootstrapping approach~\cite{conf_intervals} to compute confidence intervals for comparing SSL-based models with those trained using conventional features for each symptom. The 95\% confidence interval for the difference between group means do not contain zero, implying a  significantly better performance for SSL-based models than conventional features. We observe this behaviour for all the symptoms except for the reduced appetite for which the performance of all the models is very low.

\section{Conclusions}
In this study, we explore the application of self-supervised learning (SSL)-based speech models for detecting individual symptoms of depression. We demonstrate that SSL embeddings yield huge performance improvements compared to conventional speech features such as spectrograms, eGeMAPS, and COVAREP. Upon comparing different SSL-based speech models, we find that models encoding predominant semantic information, combined with other information, demonstrate improved performance in symptoms such as apparent and reported sadness, concentration difficulties. Meanwhile, models encoding speaker and prosodic information excel in symptoms such as inability to feel, pessimistic thoughts, and suicidal tendencies. Importantly, combining multiple SSL-based embeddings that encode various speech aspects, enhances detection performance across most depressive symptoms, signifying the importance of semantic, speaker and prosodic information. Additionally, we show that multi-task learning, while being efficient, performs in-par or better than single-task learning.

\section{Acknowledgements} 
This work has been supported by the Canada Research Chairs Program (file number 950 - 233141) and the Canadian Institutes of Health Research. We thank the Canadian Institute for Advanced Research (CIFAR) for their support. Resources used in preparing this research were provided, in part, by NSERC, the Province of Ontario, the Government of Canada through CIFAR, and companies sponsoring the Vector Institute \url{www.vectorinstitute.ai/#partners}.

\bibliographystyle{IEEEtran}
\bibliography{references.bib}
\end{document}